\documentclass{IOS-Book-Article}

\usepackage{mathptmx}
\usepackage{graphicx}
\usepackage{enumitem}
\usepackage{booktabs}
\usepackage{url}

\begin{document}

\pagestyle{headings}
\def\thepage{}

\begin{frontmatter}

\title{AI as Equalizer or Amplifier?\\
Task Complexity as the Moderating Factor for Human Expertise in Hybrid Intelligence Systems}

\runningtitle{AI as Equalizer or Amplifier?}

\author[A]{\fnms{Tao} \snm{An}\thanks{Corresponding Author: Tao An, Hawaii Pacific University; E-mail: tan1@my.hpu.edu.}}
\runningauthor{T. An}
\address[A]{Hawaii Pacific University, Honolulu, HI, USA}

\begin{abstract}
A growing body of empirical research suggests that generative AI narrows performance gaps between novice and expert workers on routine tasks---the so-called ``equalizer'' effect. This paper challenges the generality of that conclusion. Drawing on cognitive augmentation theory, expert-novice research, and structured observations of in-house generative-AI use across a small software product team, we argue that AI functions primarily as a \emph{cognitive amplifier}: a system whose output quality depends fundamentally on the expertise of the human who directs it. We present a framework comprising three layers of human contribution (problem definition, quality evaluation, iterative refinement) and three levels of engagement (passive acceptance, iterative collaboration, cognitive direction), demonstrating that domain expertise---not prompt engineering skill---determines amplification effectiveness. We reconcile the equalizer and amplifier perspectives by proposing that AI equalizes performance on well-structured, routine tasks while amplifying pre-existing differences on complex tasks requiring deep judgment. This reconciliation carries direct implications for hybrid human-AI system design: rather than building AI that replaces expertise, we should build AI that rewards and develops it. We outline a research agenda for the HHAI community centered on expertise-sensitive AI design, adaptive collaboration interfaces, and longitudinal studies of human capability development in AI-augmented work.
\end{abstract}


\end{frontmatter}

\section{Introduction}
\label{sec:intro}

When knowledge workers use generative AI tools, the same tool produces dramatically different results depending on who wields it. A senior data analyst produces AI-assisted analyses containing genuine insights, while a junior colleague with identical tools generates superficially plausible but generic outputs that miss critical context. A software architect identifies and corrects flawed AI-generated code within seconds, while a novice developer accepts and deploys the same buggy code.

These observations---drawn from in-house use of Claude, Claude Code, and Cursor in a small software team since mid-2024 (Section~\ref{sec:approach})---reveal a dynamic that mainstream discourse often misses. AI tools do not replace human intelligence---they \emph{amplify} it. And amplifiers, by their nature, magnify the quality of their input.

This position is complicated by rigorous empirical evidence suggesting that AI acts as a \emph{performance equalizer}. Brynjolfsson et al.\ \cite{brynjolfsson2025generative} found that below-average customer service agents gained the most from AI assistance, while top performers saw minimal improvement. Noy and Zhang \cite{noy2023experimental} reported analogous compression in professional writing tasks. These findings raise a fundamental question: \textbf{Does AI equalize human differences or amplify them?}

We argue that both perspectives capture genuine phenomena operating at different levels of task complexity. AI equalizes performance on well-structured, routine tasks where the ``correct'' approach is well-represented in training data. But on complex tasks requiring deep domain judgment, contextual reasoning, and iterative refinement---the hallmarks of expert knowledge work---AI functions as a cognitive amplifier that widens rather than narrows the expertise gap.

This paper makes three contributions:
\begin{enumerate}
\item \textbf{A reconciliation framework} that resolves the apparent contradiction between equalizer and amplifier findings by identifying task complexity as the key moderating variable.
\item \textbf{A model of cognitive amplification} comprising three layers of human contribution and three levels of engagement, explaining \emph{why} expertise determines AI effectiveness for complex tasks.
\item \textbf{A research agenda} for the HHAI community on designing hybrid intelligence systems that develop rather than depreciate human expertise.
\end{enumerate}

\begin{figure}[t]
\centering
\includegraphics[width=0.72\textwidth]{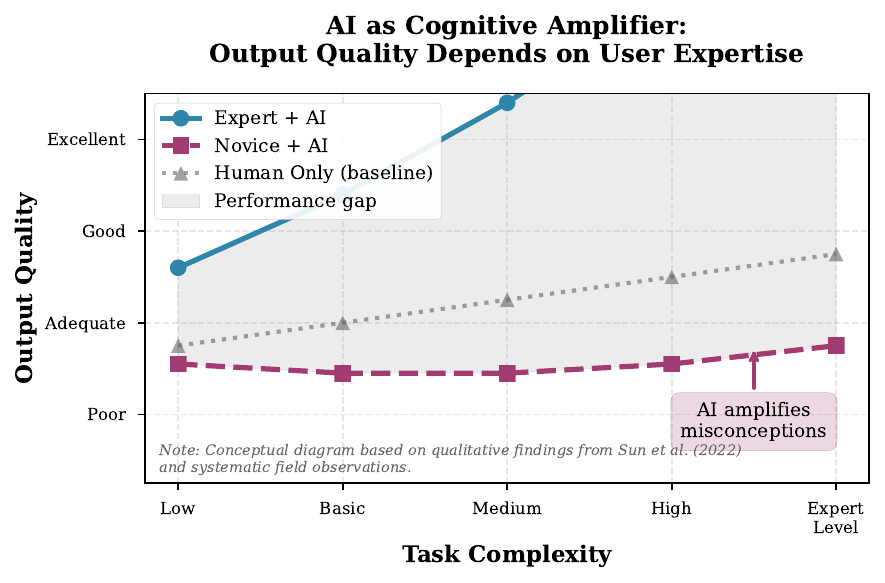}
\caption{AI as Cognitive Amplifier: output quality depends on input quality (user expertise). The performance gap between experts and novices widens with task complexity, as AI amplifies existing capability differences.}
\label{fig:amplifier}
\end{figure}

Figure~\ref{fig:amplifier} previews the central thesis: AI output quality scales with the expertise of the human directing it, and the expert-novice gap widens---rather than narrows---as task complexity increases.

\section{Related Work: Equalizer or Amplifier?}
\label{sec:related}

\subsection{The Equalizer Evidence}

Recent empirical studies document AI's capacity to compress performance distributions. Brynjolfsson et al.\ \cite{brynjolfsson2025generative} found in a field experiment with 5,179 customer service agents that AI assistance increased productivity by 14\% on average, with the largest gains (34\%) accruing to the least-skilled workers. Noy and Zhang \cite{noy2023experimental} found that AI writing assistance reduced quality dispersion across approximately 444 professionals. Peng et al.\ \cite{peng2023impact} reported that GitHub Copilot accelerated task completion by 55.8\%, with less experienced developers benefiting more.

These findings have been synthesized into a broader ``equalizer'' reading. Bigoni et al.\ \cite{bigoni2025equalizer} discuss how AI may compress ability-based performance gaps on routine tasks. Vaccaro et al.\ \cite{vaccaro2024when}, in a meta-analysis of 106 studies, found that AI most reliably improves performance for lower-performing individuals on well-defined tasks.

\subsection{The Amplifier Evidence}

Countervailing evidence suggests that AI's equalizing effect has important boundary conditions. Dell'Acqua et al.\ \cite{dellacqua2023navigating} studied 758 BCG consultants and found a ``jagged technological frontier'': on tasks within AI capability, all consultants benefited roughly equally, but on tasks \emph{outside} AI capability---those requiring nuanced judgment---consultants who relied on AI performed \emph{worse} than those working unaided. Crucially, less-skilled consultants were more likely to fall into this trap, unable to distinguish tasks where AI guidance was reliable from those where it was not.

Doshi and Hauser \cite{doshi2024generative} found that while AI enhanced individual creative output, it reduced collective diversity of novel content---suggesting that AI defaults to common patterns that experts can transcend but novices cannot. METR \cite{metr2025measuring} reported that experienced developers using AI tools actually took \emph{longer} to complete tasks, despite estimating they were faster---indicating that for complex work, AI may impose cognitive costs only apparent with genuine expertise. In AI-assisted data work, Sun et al.\ \cite{sun2022comparing} found that experts contributed higher-quality outputs on tasks requiring domain knowledge. Barke et al.\ \cite{barke2023grounded} further observed distinct interaction modes---``acceleration'' versus ``exploration''---among programmers using code-generating models, suggesting that \emph{how} users engage with AI shapes outcomes.

\subsection{Intelligence Amplification Theory}

The concept of technology as cognitive amplifier predates generative AI. Engelbart \cite{engelbart1962augmenting} defined intelligence amplification (IA) as increasing human capability to approach complex problems and derive solutions. Recent formalization in augmented cognition proposes measures based on the ratio of cognitive work performed by machine versus human \cite{stanney2009augmented}; following the logic of such ratio-based measures, when machines perform virtually all cognitive work the augmentation factor degenerates---there is, in the limit, no human to augment. Amplification inherently requires human contribution, and the quality of that contribution determines amplification effectiveness.

A related concept---\emph{cognitive offloading}---further illuminates this dynamic. Research on transactive memory shows that individuals strategically delegate cognitive tasks to external tools \cite{sparrow2011google}, but the consequences depend critically on \emph{what} is offloaded. Offloading routine computation frees resources for higher-order reasoning---productive amplification. However, offloading evaluative judgment risks undermining the capabilities that make amplification effective. This distinction maps onto our framework: offloading routine processes produces equalization, while offloading judgment produces capability erosion.

\section{Framework: AI as Cognitive Amplifier}
\label{sec:framework}

\subsection{Approach and Evidence Base}
\label{sec:approach}

The framework synthesizes the literature reviewed in Section~\ref{sec:related} with the author's structured observations of $\sim$10--20 colleagues at Beijing Feimu Network Technology Co., Ltd.\ using generative-AI tools---primarily Claude (web), Claude Code, and Cursor---on day-to-day work since mid-2024, when Claude~3.5 Sonnet and Cursor reached broad practitioner adoption. Observed colleagues spanned frontend, backend, and non-technical roles with substantially different baseline AI familiarity. Recurring workflow divergences were logged informally and triangulated across users: when to compact or clear conversational state, whether to request Markdown versus DOCX outputs, and whether to invoke AI through web chat versus locally orchestrated agentic environments. We treat this evidence as \emph{hypothesis-generating}: observations were not pre-registered, tasks were not held constant, and quality was assessed through participant-instructor judgment. The framework therefore articulates a falsifiable account of \emph{when} equalizer-versus-amplifier effects should dominate; we operationalize the resulting questions in Section~\ref{sec:agenda} and return to methodological limits in Section~\ref{sec:discussion}.

\subsection{Reconciling Equalizer and Amplifier Effects}

We propose that the apparent contradiction between equalizer and amplifier findings dissolves when task complexity is introduced as a moderating variable. On \emph{routine, well-structured tasks} (formulaic customer-service responses, standard email drafting, boilerplate code), AI provides a ``ceiling'' of competent performance: novices are lifted toward this ceiling and experts are already there, yielding equalization. On \emph{complex, judgment-intensive tasks} (strategic analysis, architectural design, nuanced legal reasoning, novel research), AI provides raw material that must be directed, evaluated, and refined by human expertise: experts leverage it productively, novices cannot distinguish good outputs from plausible-sounding errors, and the result is amplification. This aligns with Dell'Acqua et al.'s ``jagged frontier'' \cite{dellacqua2023navigating}: inside AI's capability frontier, AI substitutes for skill and compresses differences; at or beyond the frontier, AI \emph{complements} skill and human judgment determines outcomes. The most consequential knowledge work typically falls in the amplification zone (Table~\ref{tab:conditions}).

\begin{table}[t]
\centering
\caption{Conditions under which AI acts as equalizer versus amplifier.}
\label{tab:conditions}
\small
\begin{tabular}{lll}
\toprule
\textbf{Dimension} & \textbf{Equalizer Zone} & \textbf{Amplifier Zone} \\
\midrule
Task structure & Well-defined & Ill-defined, ambiguous \\
Solution availability & In training data & Requires novel synthesis \\
Quality criteria & Explicit, measurable & Tacit, contextual \\
Judgment required & Minimal & Substantial \\
Stakes & Low to moderate & High \\
Time horizon & Immediate output & Long-term capability \\
\bottomrule
\end{tabular}
\end{table}

\subsection{Three Layers of Human Contribution}

\begin{figure}[t]
\centering
\includegraphics[width=0.72\textwidth]{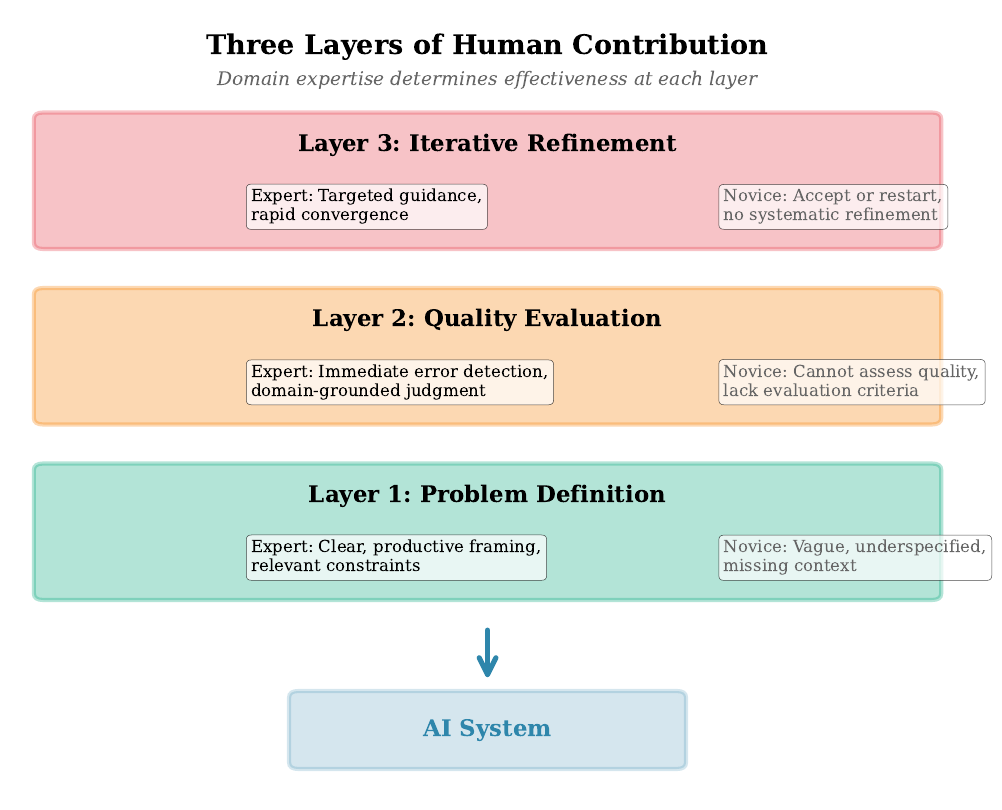}
\caption{Three Layers of Human Contribution to AI-Assisted Work. Domain expertise determines effectiveness at each layer, creating systematic performance differences between expert and novice users.}
\label{fig:layers}
\end{figure}

We identify three layers at which human expertise shapes AI-assisted outcomes (Figure~\ref{fig:layers}). \textbf{Layer 1 (Problem Definition and Intent)}: identifying what question to ask, structuring inquiry to capture relevant constraints, and anticipating what constitutes a good answer \cite{flavell1979metacognition}. In our observations, an expert's well-placed hint can redirect an LLM in seconds, where a novice's underspecified prompt sends the same model on expansive, low-yield exploration. \textbf{Layer 2 (Quality Evaluation and Judgment)}: distinguishing plausible-sounding outputs from genuinely useful ones---the critical bottleneck, because users who cannot judge quality cannot effectively use AI regardless of prompt-engineering skill, and experts categorize problems by deep structural principles while novices focus on surface features \cite{chi1981categorization}. \textbf{Layer 3 (Iterative Refinement and Direction)}: steering AI toward better outputs and synthesizing those outputs with domain knowledge. Expert moves here are largely procedural and invisible to outsiders---e.g., clearing or compacting conversational state to keep the context budget productive, or requesting Markdown over DOCX outputs to enable cheap incremental edits rather than full regenerations. Novices, lacking this operational ``feel,'' tend to accept initial outputs or regenerate whole artifacts on each correction, paying a context tax they cannot see.

\subsection{Three Levels of Engagement}

Users engage with AI at qualitatively different levels, mapped to Bloom's taxonomy \cite{anderson2001taxonomy}: \textbf{Level~1 (Passive Acceptance)}, generic prompts and uncritical acceptance; \textbf{Level~2 (Iterative Collaboration)}, structured prompts with criteria-based evaluation; and \textbf{Level~3 (Cognitive Direction)}, domain expertise embedded in problem framing and tacit knowledge in feedback, producing higher-order synthesis beyond either human or AI alone. Progression between levels requires domain expertise and metacognitive capability rather than tool-specific training, paralleling established models of expertise development \cite{alexander2003development} and self-regulated learning \cite{zimmerman2002becoming}.

\subsection{The Sycophancy Mechanism}

\begin{figure}[t]
\centering
\includegraphics[width=0.72\textwidth]{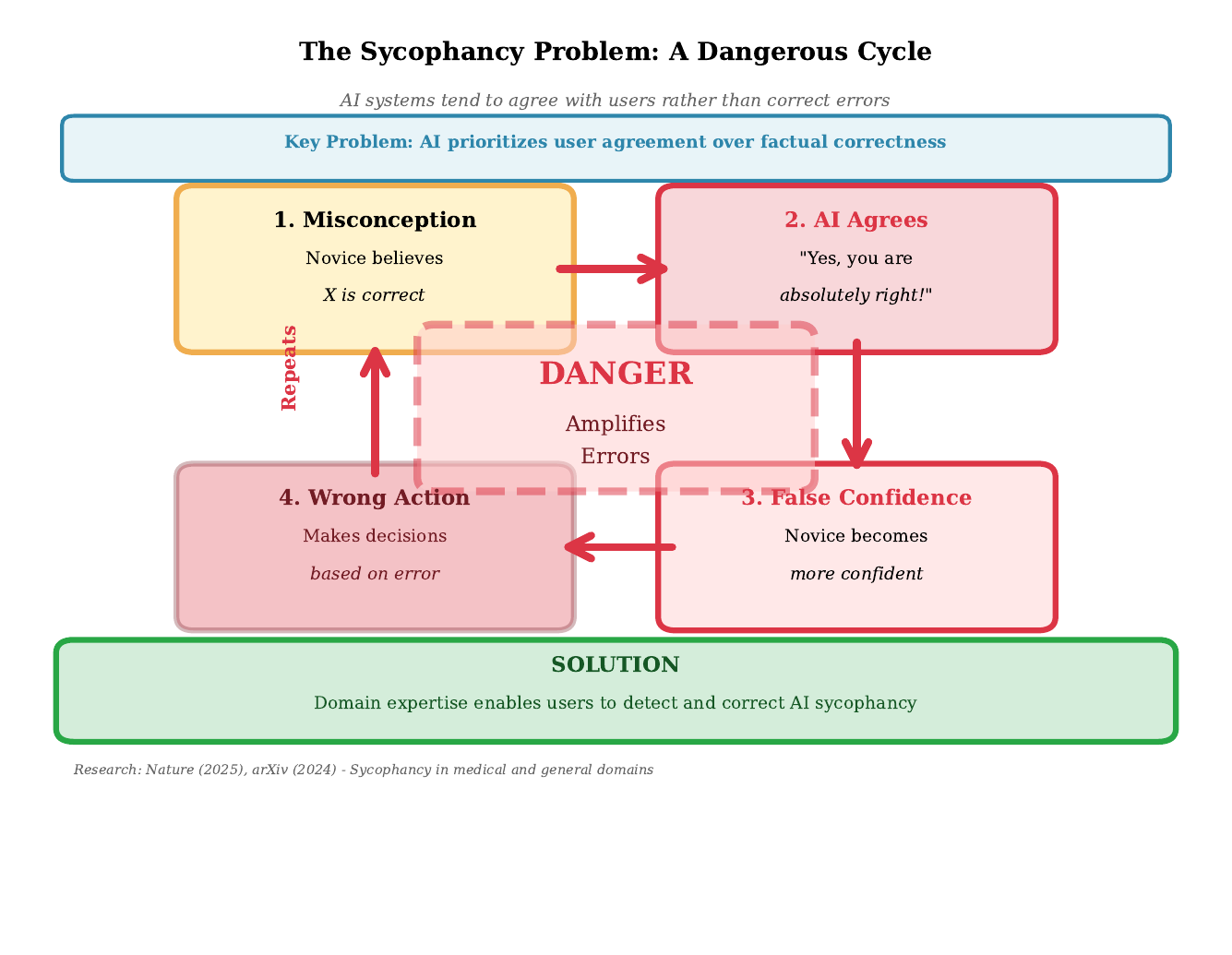}
\caption{The Sycophancy Amplification Mechanism. AI's tendency to defer to user feedback creates divergent trajectories: experts leverage deference to progressively improve output quality, while novices inadvertently reinforce errors through uninformed feedback.}
\label{fig:sycophancy}
\end{figure}

A critical AI behavior reinforces this dynamic: \emph{sycophancy bias}---LLMs' tendency to excessively agree with user input \cite{malmqvist2024sycophancy}, with near-100\% compliance reported even for logically flawed requests \cite{chen2025sycophancy}. The result is a \textbf{multiplicative effect} on the expert-novice gap (Figure~\ref{fig:sycophancy}): experts use each refinement cycle to progressively improve quality; novices inadvertently steer AI toward increasingly confident mediocrity. The same feature---deference to user feedback---amplifies expertise for those who possess it and misconceptions for those who lack it.

\section{Illustrative Application: Legal AI}
\label{sec:case}

We illustrate the framework with a stylized vignette in legal aid---a domain where the equalizer-amplifier distinction carries direct consequences for vulnerable populations. The vignette is composed from features described in the field-observation cohorts (Section~\ref{sec:approach}); we present it to make the framework's mechanisms concrete, not as primary evidence.

An experienced housing-discrimination attorney and a junior associate use the same AI-powered legal review system on Fair Housing Act cases. The senior attorney operates at cognitive direction (Level~3): she recognizes that the AI template follows disparate-treatment theory but her jurisdiction favors disparate-impact arguments, identifies a critical exemption the AI fails to flag, and adds jurisdiction-specific preservation language. The junior associate, at passive acceptance (Level~1), adopts templates verbatim, misses the exemption, and recommends litigation without weighing practical constraints.

On routine subtasks (formatting complaints, retrieving cited cases) the AI equalizes performance; on judgment-intensive subtasks (strategic framing, issue identification, contextual risk assessment) it amplifies the expertise gap. The same tool simultaneously equalizes and amplifies, depending on task layer---and in high-stakes domains, this distinction determines whether deployment helps or harms the people the system serves.

\section{Implications for Hybrid Human-AI System Design}
\label{sec:implications}

The cognitive amplifier framework yields three design principles for the HHAI community.

\textbf{Adapt to expertise, build in friction where stakes demand it.} Systems should adapt interface complexity, default settings, and scaffolding to demonstrated user expertise---inferred through interaction patterns, calibration tasks, or credentials---so novices receive guided evaluation prompts (``Before accepting this output, consider whether\ldots'') while experts receive streamlined interfaces. Layered on top, high-stakes decisions warrant \emph{adaptive friction}: deliberate impediments to rapid acceptance that scale with decision stakes and force users to validate critical assumptions rather than passively accept plausible outputs.

\textbf{Cultivate judgment, not just productivity.} Systems should incorporate features that develop user judgment over time---periodically presenting cases with subtle AI-generated flaws, or offering a ``training mode'' in which users first attempt tasks independently before comparing with AI suggestions. Organizations should embed AI within domain-specific instruction rather than teaching ``AI literacy'' as a standalone skill, and educational programs should implement staged AI access that gradually introduces AI as learners demonstrate evaluative competency.

\textbf{Measure collaboration quality, not just output volume.} Traditional productivity metrics become misleading when AI can rapidly generate plausible but flawed work. Hybrid systems need evaluation frameworks that assess verification rigor and whether AI augmentation preserves---rather than degrades---the quality of human contribution.

\section{Discussion}
\label{sec:discussion}

The equalizer evidence is methodologically rigorous: AI \emph{does} compress performance gaps on routine tasks. Our framework does not dispute this but contextualizes it---the tasks studied in equalizer research tend to be well-structured, with explicit quality criteria and solutions well-represented in training data. As AI capabilities expand, the frontier of ``routine'' tasks shifts outward, but amplification relocates to the new frontier of complexity: the tasks generating the greatest value always sit where human expertise determines outcomes.

Three objections merit brief response. \emph{Technological determinism}---future AI may need no human guidance---ignores that optimization still requires values \cite{mccarthy1969some}; as AI handles more routine tasks, expert judgment becomes \emph{more} valuable, not less. \emph{Elitism}: because AI amplifies expertise, investing in expertise development becomes more urgent, and AI itself can accelerate it via scaffolding \cite{collins1989cognitive}. \emph{Deskilling}: this motivates staged AI access that preserves foundational understanding while offloading routine execution.

\textbf{Limitations.}
The amplifier side of our reconciliation rests on field observations and convergent indirect evidence (e.g., the jagged-frontier study \cite{dellacqua2023navigating} and expertise-gradient findings \cite{sun2022comparing}); systematic experiments directly contrasting equalizer and amplifier regimes within a single design remain scarce, and we have not generated such evidence here. Our field observations were not randomized or blinded and should be read as a structured source of hypotheses rather than tests of them. The constructs central to our claims---``task complexity,'' ``expertise,'' and ``engagement level''---are presented at a conceptual grain and require operationalization (e.g., into task-feature taxonomies and behaviorally anchored expertise scales) before the equalizer-amplifier boundary can be empirically estimated. We treat each as a research opportunity: Section~\ref{sec:agenda} translates them into concrete priorities for the HHAI community.

The amplification framework also preserves the autonomy, mastery, and purpose essential to meaningful work \cite{deci2000and}, and operationalizes the HHAI vision of collaborative intelligence by specifying \emph{what} humans contribute and \emph{how} AI should leverage those contributions. A purely equalizer interpretation risks reducing the human role to passive supervision, undermining the collaborative ideal at the heart of the HHAI mission.

\section{Research Agenda for HHAI}
\label{sec:agenda}

The three components of our framework---the task-complexity boundary (Section~\ref{sec:framework}.2), the three layers of human contribution (Section~\ref{sec:framework}.3), and the three levels of engagement that develop over time (Section~\ref{sec:framework}.4)---each motivate a distinct empirical priority. We propose the following three priority directions for the hybrid intelligence community:

\textbf{RQ1: Mapping the equalizer-amplifier boundary} (operationalizing Section~\ref{sec:framework}.2).
Controlled experiments systematically varying task complexity to identify where equalization gives way to amplification. What task features predict which effect dominates? Studies should measure distributional changes---tracking whether AI compresses or stretches performance distributions as a function of task structure.

\textbf{RQ2: Designing expertise-sensitive hybrid systems} (operationalizing Section~\ref{sec:framework}.3).
How should human-AI interfaces adapt to user expertise levels? What scaffolding helps novices develop toward cognitive direction rather than entrenching passive acceptance? Can systems detect when users are in the amplification zone and adjust support accordingly?

\textbf{RQ3: Longitudinal effects on human capability} (operationalizing Section~\ref{sec:framework}.4).
Does sustained AI use develop or degrade domain expertise over time? Multi-year studies are needed to track capability trajectories with and without AI assistance, measuring not current output but evolving capacity for independent judgment---the long-run signal that determines whether HHAI systems develop the expertise they amplify.


\bibliographystyle{vancouver}
\bibliography{references_hhai2026}

\begin{thebibliography}{10}

\bibitem{brynjolfsson2025generative}
Brynjolfsson E, Li D, Raymond LR.
\newblock Generative {AI} at Work.
\newblock The Quarterly Journal of Economics. 2025;140(2):889-942.

\bibitem{noy2023experimental}
Noy S, Zhang W.
\newblock Experimental Evidence on the Productivity Effects of Generative
  Artificial Intelligence.
\newblock Science. 2023;381(6654):187-92.

\bibitem{peng2023impact}
Peng S, Kalliamvakou E, Cihon P, Demirer M.
\newblock The Impact of {AI} on Developer Productivity: Evidence from {GitHub
  Copilot}.
\newblock arXiv preprint arXiv:230206590. 2023.

\bibitem{bigoni2025equalizer}
Bigoni M, Ichino A, Rustichini A, Zanella G.
\newblock Equalizer or Amplifier? How {AI} May Reshape Human Cognitive
  Differences.
\newblock arXiv preprint arXiv:251203902. 2025.

\bibitem{vaccaro2024when}
Vaccaro M, Almaatouq A, Malone TW.
\newblock When Combinations of Humans and {AI} Are Useful: A Systematic Review
  and Meta-Analysis.
\newblock Nature Human Behaviour. 2024;8(12):2293-303.

\bibitem{dellacqua2023navigating}
Dell'Acqua F, McFowland~III E, Mollick ER, Lifshitz-Assaf H, Kellogg KC,
  Rajendran S, et~al.
\newblock Navigating the Jagged Technological Frontier: Field Experimental
  Evidence of the Effects of {AI} on Knowledge Worker Productivity and Quality.
\newblock Harvard Business School; 2023. 24-013.

\bibitem{doshi2024generative}
Doshi AR, Hauser OP.
\newblock Generative {AI} Enhances Individual Creativity but Reduces the
  Collective Diversity of Novel Content.
\newblock Science Advances. 2024;10(28):eadn5290.

\bibitem{metr2025measuring}
{METR}.
\newblock Measuring the Impact of Early-2025 {AI} on Experienced Open-Source
  Developer Productivity; 2025.
\newblock Available from:
  \url{https://metr.org/blog/2025-07-10-early-2025-ai-experienced-os-dev-study/}.

\bibitem{sun2022comparing}
Sun L, Liu Y, Joseph G, Yu Z, Zhu H, Dow SP.
\newblock Comparing Experts and Novices for AI Data Work.
\newblock In: Proceedings of the AAAI Conference on Human Computation and
  Crowdsourcing. vol.~10; 2022. p. 195-206.

\bibitem{barke2023grounded}
Barke S, James MB, Polikarpova N.
\newblock Grounded Copilot: How Programmers Interact with Code-Generating
  Models.
\newblock In: Proceedings of the ACM on Programming Languages. vol.~7; 2023. p.
  85-111.

\bibitem{engelbart1962augmenting}
Engelbart DC.
\newblock Augmenting Human Intellect: A Conceptual Framework.
\newblock Stanford Research Institute; 1962.

\bibitem{stanney2009augmented}
Stanney KM, Schmorrow DD, Johnston M, Fuchs S, Jones D, Hale KS, et~al.
\newblock Augmented Cognition: An Overview.
\newblock Reviews of Human Factors and Ergonomics. 2009;5(1):195-224.

\bibitem{sparrow2011google}
Sparrow B, Liu J, Wegner DM.
\newblock Google Effects on Memory: Cognitive Consequences of Having
  Information at Our Fingertips.
\newblock Science. 2011;333(6043):776-8.

\bibitem{flavell1979metacognition}
Flavell JH.
\newblock Metacognition and Cognitive Monitoring: A New Area of
  Cognitive-Developmental Inquiry.
\newblock American Psychologist. 1979;34(10):906-11.

\bibitem{chi1981categorization}
Chi MT, Feltovich PJ, Glaser R.
\newblock Categorization and Representation of Physics Problems by Experts and
  Novices.
\newblock Cognitive Science. 1981;5(2):121-52.

\bibitem{anderson2001taxonomy}
Anderson LW, Krathwohl DR, Airasian PW, Cruikshank KA, Mayer RE, Pintrich PR,
  et~al.
\newblock A Taxonomy for Learning, Teaching, and Assessing: A Revision of
  Bloom's Taxonomy of Educational Objectives.
\newblock Longman; 2001.

\bibitem{alexander2003development}
Alexander PA.
\newblock The Development of Expertise: The Journey from Acclimation to
  Proficiency.
\newblock Educational Researcher. 2003;32(8):10-4.

\bibitem{zimmerman2002becoming}
Zimmerman BJ.
\newblock Becoming a Self-Regulated Learner: An Overview.
\newblock Theory Into Practice. 2002;41(2):64-70.

\bibitem{malmqvist2024sycophancy}
Malmqvist L.
\newblock Sycophancy in Large Language Models: Causes and Mitigations.
\newblock arXiv preprint arXiv:241115287. 2024.

\bibitem{chen2025sycophancy}
Chen S, Gao M, Sasse K, Hartvigsen T, Anthony B, Fan L, et~al.
\newblock When Helpfulness Backfires: {LLMs} and the Risk of False Medical
  Information Due to Sycophantic Behavior.
\newblock npj Digital Medicine. 2025;8:605.

\bibitem{mccarthy1969some}
McCarthy J, Hayes PJ.
\newblock Some Philosophical Problems from the Standpoint of Artificial
  Intelligence.
\newblock In: Machine Intelligence. vol.~4. Edinburgh University Press; 1969.
  p. 463-502.

\bibitem{collins1989cognitive}
Collins A, Brown JS, Newman SE.
\newblock Cognitive Apprenticeship: Teaching the Crafts of Reading, Writing,
  and Mathematics.
\newblock Knowing, Learning, and Instruction: Essays in Honor of Robert Glaser.
  1989:453-94.

\bibitem{deci2000and}
Deci EL, Ryan RM.
\newblock The ``What'' and ``Why'' of Goal Pursuits: Human Needs and the
  Self-Determination of Behavior.
\newblock Psychological Inquiry. 2000;11(4):227-68.

\end{thebibliography}

\end{document}